\documentclass[12pt]{iopart}
\usepackage{graphicx}
\usepackage{psfrag}
\usepackage{iopams}

\def\be{\begin{equation}}             \def\ee{\end{equation}}
\def\bea{\jot0ex\begin{eqnarray}}     \def\eea{\end{eqnarray}}

\def\kago{kagom\'{e}}
\def\ND{N_{\bigtriangledown}}
\def\CN{\mathcal{N}}
\def\SPN{$\rm Sp(\CN)$}
\def\EMF{E_{\rm MF}}

\def\la{\lambda_a}
\def\lc{\lambda_c}
\def\S{{\mathbf{S}}}
\def\k{{\mathbf{k}}}

\def\qord{{\mathbf{q}_{ord}}}
\def\kmin{{\mathbf{k}_{min}}}
\begin{document}

\title{Spatially anisotropic Heisenberg Kagom\'{e} anti\-ferro\-magnet}
\author{W. Apel$^1$, T. Yavors'kii$^2$ and H.-U. Everts$^3$}
\address{$^1$ Physikalisch-Technische Bundesanstalt, Braunschweig, Germany}
\address{$^2$ Department of Physics, University of Waterloo, Ontario, Canada}
\address{$^3$ Institut f\"{u}r Theoretische Physik, Universit\"{a}t Hannover,
Hannover, Germany}

\ead{Walter.Apel@ptb.de}

\begin{abstract}

In the search for spin-$1/2$ \kago{} antiferromagnets, the mineral
volborthite has recently been the subject of experimental studies 
\cite{hiroi_et_al01,fu_et_al03,be-et-al04,be-et-al05}. 
It has been suggested that the magnetic properties of this material are 
described by a spin $1/2$ Heisenberg model on the  kagom\'{e} lattice 
with spatially anisotropic exchange couplings. 
We report on investigations of the \SPN{} symmetric generalisation of 
this model in the large $\CN$ limit. 
We obtain a detailed description of the dependence of possible ground states 
on the anisotropy  and on the spin length $S$. 
A fairly rich phase diagram with a ferrimagnetic phase, incommensurate phases 
with and without long range order and a decoupled chain phase emerges.

\end{abstract}

\vspace*{-2ex}
\pacs{75.10.Jm, 75.30.Kz}
\vspace*{-2ex}
\submitto{\JPCM}

\section{Introduction}

\begin{figure}
\psfrag{d1}{$\bdelta_1$}
\psfrag{d2}{$\bdelta_2$}
\psfrag{d3}{$\bdelta_3$}
\psfrag{J}{$J$}
\psfrag{Js}{$J^{\prime}$}
\centerline{\includegraphics[width=9cm]{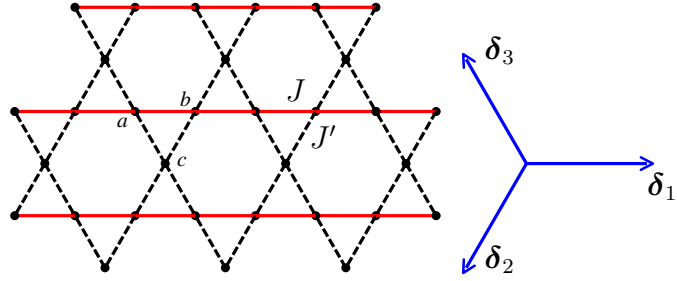}}
\caption{Anisotropic {\kago} model. 
The coupling $J'$ will be set equal to unity.
$\bdelta_{\nu}$, $\nu = 1, 2, 3$, are the primitive lattice vectors of 
the \kago{} net.}
\label{aniskago}
\end{figure}

The magnetic lattice of the natural antiferromagnet volborthite consists of 
$\rm {Cu}^{2+}$ ions which occupy well separated planar kagom\'{e}-like nets.
A monoclinic distortion leads to a slight difference
between the exchange couplings along one lattice direction ($J$) and
the two other directions ($J'$) within the the planes. In this study, we  consider the 
appropriate model, the Heisenberg antiferromagnet on the \kago{} lattice with 
spatial anisotropy  (see \fref{aniskago}), 
\be \fl \hspace*{5mm}
\mathcal{H} \;=\; \sum_{\mbox{n.n.}} J_{ij} \; \S_i \, \S_j 
\mbox{\hspace*{3mm}with\hspace*{2mm}}
 J_{ij} = \left\{\begin{array}{l@{\hspace*{5mm}}l} J &  
   \mbox{if $i$ and $j$ $\not\in$ c-sublattice} \\
         J'=1 & \mbox{otherwise} \end{array}  \right.\,.
\label{akago}
\ee

We have applied the spin wave approximation to this model, have done  quantum 
perturbation theory around trial groundstates, and have studied the \SPN{}-generalisation 
in the limit of large $\CN$ where the mean-field approximation becomes exact \cite{YAE07}.
Here, we report the results of the \SPN{}-symmetric model. We consider this model 
in the full range of the anisotropy $0 \leq J \leq \infty$ and  pay particular attention 
to possible transitions between  different phases of the model which are expected to 
occur when $J$ is varied. 

\section{\SPN{} approach}
In this section, we sketch the mean field treatment 
of the \SPN{} generalisation of the model \eref{akago}.
The procedure follows closely that of \cite{rea_sac91,sac92}. 
Further details can also be found in \cite{YAE07}.
The first step is the replacement of the $\rm{SU(2)}$ invariant product 
$\S_i \cdot \S_j$ in \eref{akago} by the \SPN{}-symmetric expression 
$ \frac{1}{\mathcal{N}} (\mathcal{J}_{\alpha \beta} b^{\dagger\,\alpha} _{i} 
b^{\dagger\,\beta }_{j}) (\mathcal{J}^{\gamma \delta} b_{i\gamma}  b_{j\delta})$,
where $b_{i\alpha}$, $b^{\dagger\,\alpha}_{i}$  are bosonic operators 
and $\mathcal{J}$ is the $\mathcal{N}$-dimensional generalisaton of the 
antisymmetric unit tensor in two dimensions \cite{rea_sac91,sac92}.
Here, we have introduced a paired index notation,
 $\alpha = (m,\sigma),\, m=1,\cdots ,\mathcal{N},\; \sigma = \uparrow, \downarrow$ 
for the spinor index of the bosonic operators. 
The boson occupation numbers $n_{b} =  b_i^{\dagger\,\alpha} \; b_{i\alpha}$
are subject to the constraint 
\be
\kappa = n_b/\mathcal{N} 
\label{constr}
\ee
with $\kappa$ a fixed parameter \cite{sac92}. For the $SU(2)$ case, $\mathcal{N} = 1$, one has 
$\kappa = 2S$ with $S$ the spin length. 
Hence, large values of  $\kappa$, equivalent to large values 
of $S$ in the $SU(2)$ model, correspond to the classical limit and  small values of $\kappa$
correspond to the quantum regime.
The next step is a decoupling of the quartic terms in the Hamiltonian 
by a Hubbard-Stratonovich transformation using fields $Q_{ij}$ which are 
defined on the links of the lattice.
Additional fields $\lambda$ have to be introduced as Lagrange multipliers to enforce 
the constraints \eref{constr} on the sites. 
We assume that the link variables are identical
on all horizontal links  of  \fref{aniskago},  $|Q_{ab}|=Q_1$, and on all diagonal links 
of \fref{aniskago}, $|Q_{ac}|=|Q_{bc}|= Q_2$. 
Furthermore, we assume that $\lambda=\la$ on the a and b sites in 
\fref{aniskago}, and  $\lambda=\lc$ on the c sites. 
Now, the bosons can be integrated out exactly. 
Then, in the large $\mathcal{N}$ limit,
the physical values of the variables $Q_1$, $Q_2$, $\la$, and $\lc$ are determined by 
 the saddlepoint of the mean field energy  
\be \fl \hspace*{5mm}
\frac{\EMF}{\CN \ND} \;=\;
 J\, |Q_1|^2 + 2\, |Q_2|^2 - (2 \la + \lc)(\kappa +1) 
 + \frac{1}{\ND} \sum_{\k,\mu}\omega_{\mu}(\k)
  \left(1 + |{x}_{\mu}(\k)|^2 \right) \label{EMFK}
\ee
with respect to these variables.
Here, $\ND$ is the number of downward pointing triangles of the lattice. 
The spinon frequencies $\omega_{\mu}(\k)$, $\mu = 1,\,2,\,3$, are obtained as 
the three positive solutions of $\det \mathbf{\hat{D}}(\omega) = 0$, where
$\mathbf{\hat{D}}(\omega)$ is the six dimensional matrix  related to the bosonic 
part of the \SPN{} Hamiltonian after the Hubbard-Stratonovich transformation has 
been performed:
\vspace*{2mm}
\be \fl \hspace*{1cm}
\mathbf{\hat{D}}(\omega) 
 = \left(\begin{array}{ccc@{\hspace*{6mm}}ccc}
\la-\omega  & 0 & 0 & 0 & \tilde{Q}_2(\k) & J \tilde{Q}_1(\k) \\[2mm]
0 & \lc-\omega  & 0 &  \tilde{Q}_2(\k) & 0 & \tilde{Q}_3(\k) \\[2mm]
0 & 0 & \la-\omega  & J \tilde{Q}_1(\k) & \tilde{Q}_3(\k) & 0 \\[4mm]
 0 & \tilde{Q}^*_2(\k) & J \tilde{Q}^*_1(\k) & \la+\omega  & 0 & 0  \\[2mm]
\tilde{Q}^*_2(\k) & 0 & \tilde{Q}^*_3(\k) & 0 & \lc+\omega  & 0  \\[2mm]
J \tilde{Q}^*_1(\k) & \tilde{Q}^*_3(\k) & 0 & 0 & 0 & \la+\omega  
\end{array} \right)\,,
\label{DP}
\ee
\vspace*{2mm}
with  
$ 
 \tilde{Q}_{\mu}(\k)  =  i Q_{\mu} \sin({\bdelta}_{\mu} \k /2), 
\; {\mu}=1,\,2,\,3$, and $ Q_3 = Q_2. 
$
In the above formula, the matrix elements $\tilde{Q}_{\mu}(\k)$ incorporate
information on the lattice symmetries via the primitive lattice vectors 
$\bdelta_{\mu}$ of the \kago{} net, see \fref{aniskago}, and the wave vector $\k$. 
The quantity $|{x}_{\mu}(\k)|^2$ represents the density of  Bose condensate 
that will develop at those points in the Brillouin zone where one of the 
spinon frequencies vanishes. 
By calculating the \SPN{} analog of the spin-spin correlation function, 
we can show that the long distance behaviour of this function 
is directly related to the existence of a condensate \cite{YAE07}: a finite condensate density, 
$|{x}_{\mu}(\k)|^2\, > \,0$  implies long-range order (LRO), while for  
$|{x}_{\mu}(\k)|^2\,= \,0$ there is only short-range order (SRO).

To find  the saddle point of $E_{\rm{MF}}$,  
\eref{EMFK}, in the space of the variables $Q_1$, $Q_2$, 
$\lambda_a$, $\lambda_c$ and $|{x}_{\mu}(\k)|^2$ at a given pair of parameters $J$, $\kappa$, 
we use an iterative numerical routine. 
Remarkably, and in contrast to previous applications of the mean field 
\SPN{} approach \cite{rea_sac91, sac92, chu_mar_mck01}, 
it turns out to be indispensable 
to allow for two independent Lagrange multipliers $\lambda_a$ and  $\lambda_c$. In view of 
the anisotropy of our model this is plausible. 
\section{ Results and discussion}
\begin{figure}[t]
(a) \hspace{6.5cm}(b)

\includegraphics[width=6.5cm]{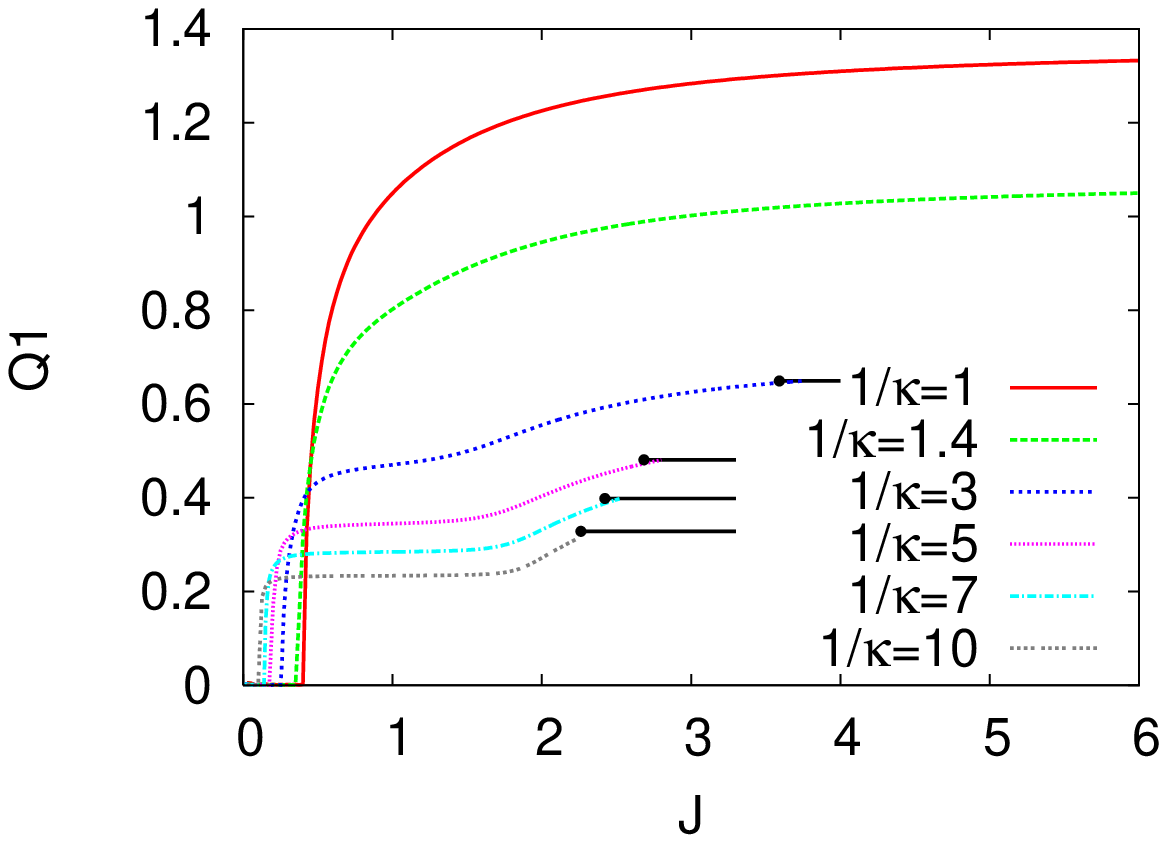}
\includegraphics[width=6.5cm]{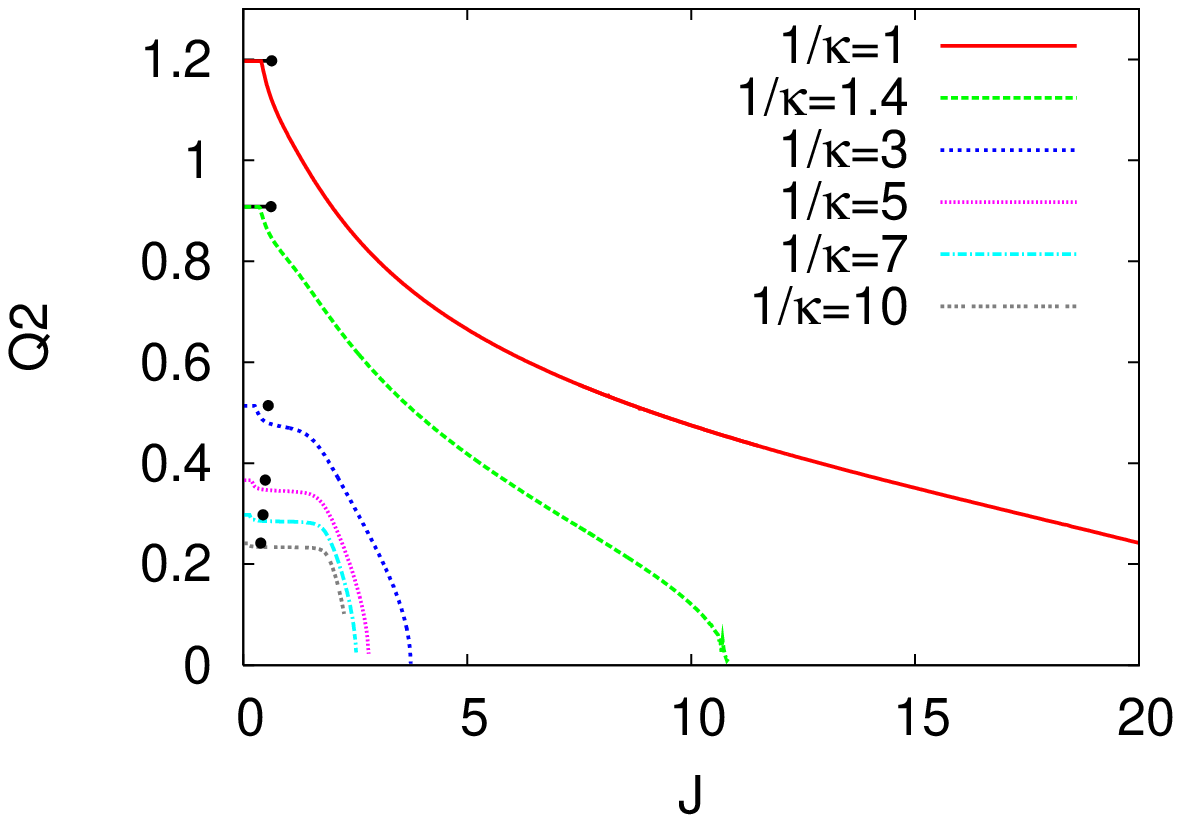}

(c) \hspace{6.5cm}(d)

\includegraphics[width=6.5cm]{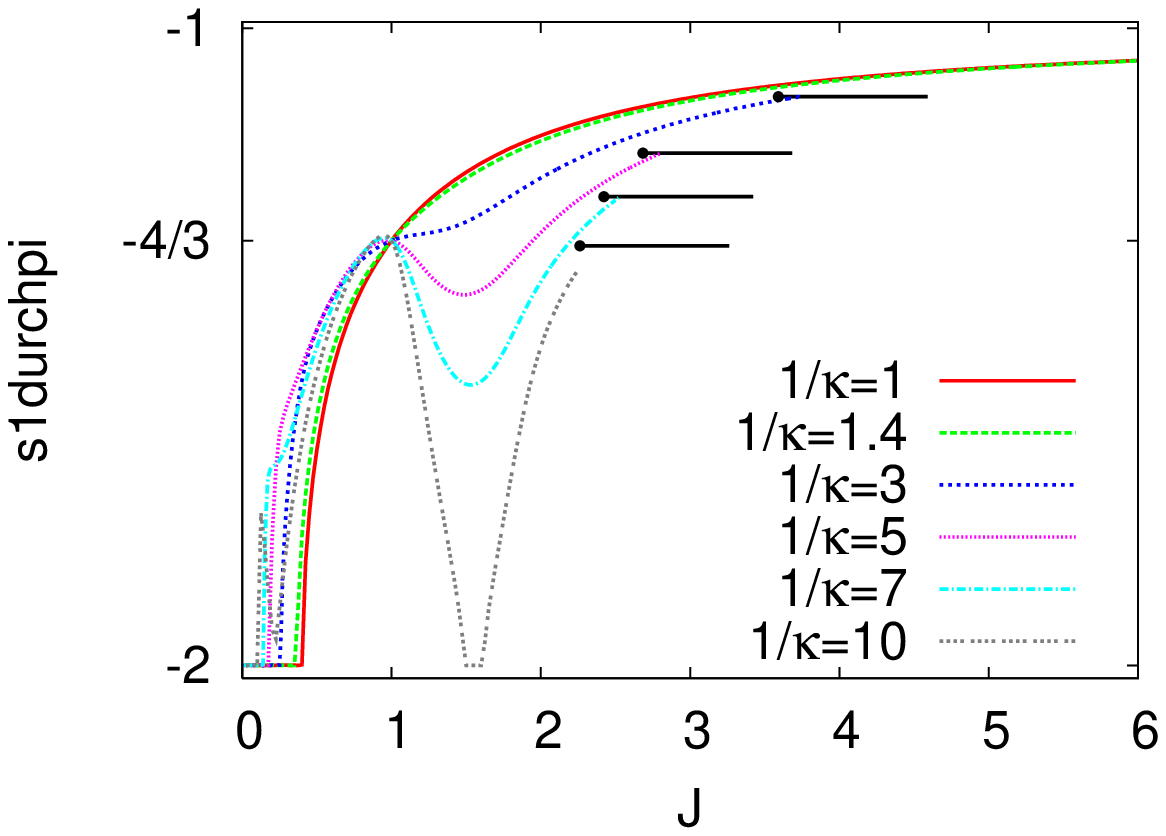}
\includegraphics[width=6.5cm]{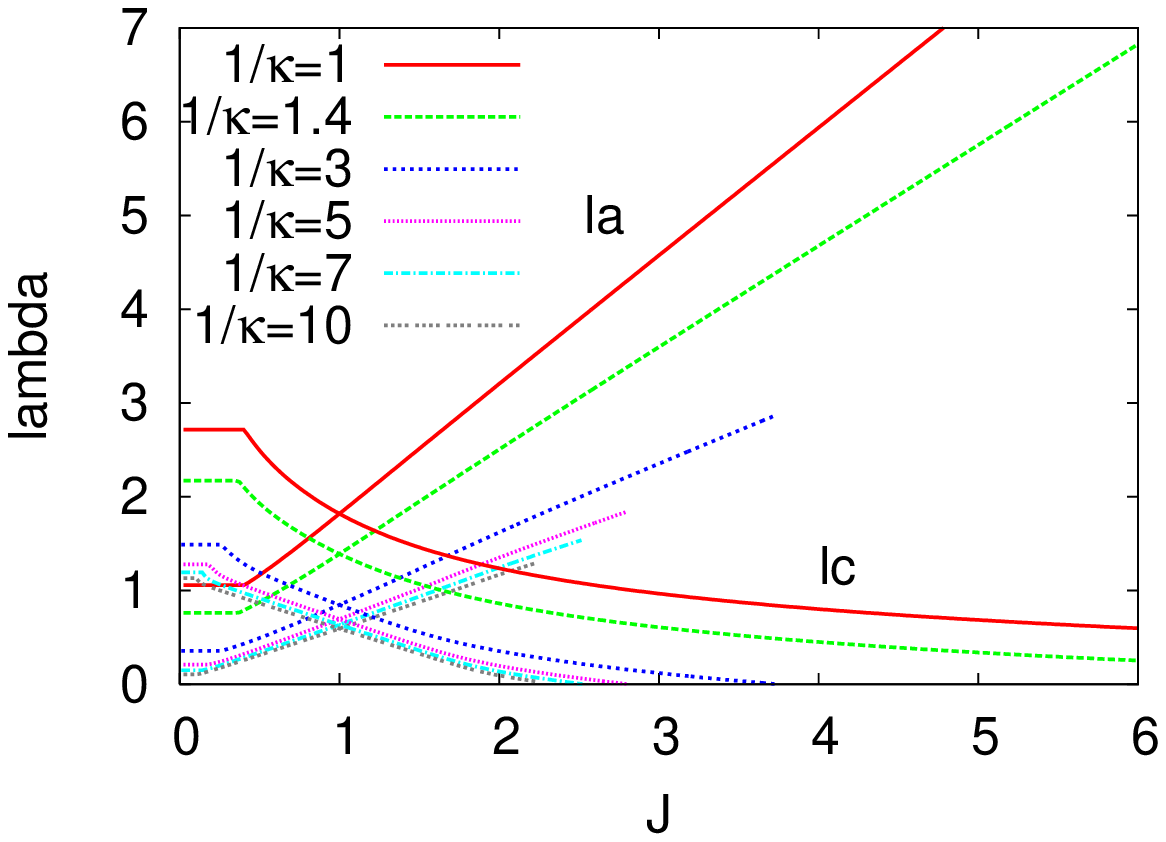}

\caption{Saddlepoint values of the parameters $Q_1$, $Q_2$, 
of the ordering wave number $q^x_{ord}$ 
($q^y_{ord}=0$), and of $\la$ and $\lc$.
Full circles and horizontal lines in panels (a), (b), and (c): 
results of analytical determination of the saddlepoint.} 
\label{results}
\end{figure}

The results of the large-$\CN$ approach are summarised in the zero 
temperature phase diagram of the anisotropic {\kago} antiferromagnet.
\psfrag{J}{$J$}
\psfrag{1/kappa}{$1/\kappa$}
\psfrag{FM}{FM}
\psfrag{DC}{DC}
\psfrag{SRO}{\parbox{4em}{IC\\SRO}}
\psfrag{LRO}{\parbox{4em}{IC\\LRO}}
\begin{figure}[h]
\centerline{\includegraphics[width=8cm]{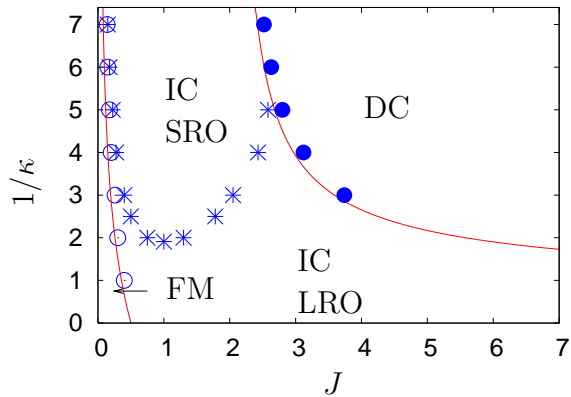}}
\caption{Phase diagram; FM: ferrimagnetic phase, IC: incommensurate phase,
DC: decoupled-chain phase; 
empty circles, full circles, and asterisks: numerical results;
full lines: analytical results. 
Asterisks mark the boundary between phases with long-range order (LRO) and 
short-range order (SRO).}
\label{pd}
\end{figure}
It displays the phases that we were able to discern in the 
$J\,-\,1/\kappa$ plane.  
We first note that as expected LRO which is always 
present for large $\kappa$ disappears from the incommensurate (IC) phase 
for sufficiently small values of $\kappa$. Recall that for 
large $\kappa$ we are in the classical regime of our model.
The phase boundary that separates the region with SRO 
from the region with LRO was found by determining for each value of $J$ 
that value of $\kappa$ where the gap in the lowest branch of the one 
spinon spectrum $\omega_{\mu}({\mathbf k})$ first vanishes as ${\kappa}$ 
is increased. 
At larger values of $\kappa$,  condensate, and hence LRO will appear. 
Since our model is maximally frustrated for $J=1$, one expects 
a maximal suppression of LRO by quantum effects at this point. 
As is seen in \fref{pd}, this is confirmed in the large-$\CN$ 
approach. 
For $J=0$, the exact quantum ground state of our model \eref{akago} is 
ferrimagnetic (FM) according to the Lieb-Mattis theorem \cite{liebma92}. 
The classical ferrimagnet has all spins on the horizontal lattice lines 
aligned ferromagnetically while the middle spins are oriented 
antiparallel to these. 
In the classical limit, $\kappa \rightarrow \infty$, of our \SPN{}-model, the expectation value 
$Q_1 = \langle \mathcal{J}^{m\sigma\, m'\sigma'} b^{\dagger}_{a\,m\sigma} 
 b^{\dagger}_{b\, m'\sigma'} ) \rangle$, which measures the singlet weight on 
the horizontal bonds, should vanish in this state for sufficiently small 
$J$, $J\leq J_{\rm{FM}}(\kappa)$.
As $J$ is increased beyond $J_{\rm{FM}}$, $Q_1$ increases in the manner of 
an order parameter at a second order phase transition 
(see panel (a) of \fref{results}). 
At the same time, the parameter $Q_2$ begins to decrease, and it drops 
to zero at  $J = J_{\rm{DC}}(\kappa)$ (see panel (b) of \fref{results}).
(Note that the $J$-scales are different in  panel (a) and panel (b)).
Thus, the  large-$\CN$ approach predicts the existence of a 
decoupled--chain phase in the region above the phase boundary 
$J_{\rm{DC}}$. 
\psfrag{J}{$J$}
\psfrag{Q1}{$Q_1$}
\psfrag{Q2}{$Q_2$}
\psfrag{lambda}{$\lambda$}
\psfrag{la}{$\la$}
\psfrag{lc}{$\lc$}
\psfrag{s1durchpi}{$q^x_{ord}/\pi$}
\psfrag{condensate}{condensate}
$Q_2$  decreases to zero continuously so that the phase transition at 
$J_{\rm{DC}}$ appears to be of second order again.
Both, LRO and SRO phases may be characterised by an ordering wave vector 
$\qord = 2\kmin$, where $\kmin$ is that wave vector at which the lowest 
branch of the one-spinon excitation spectrum $\omega_{\mu}(\k)$ 
has its minimum. 
The static spin structure factor $S(\mathbf q)$  develops a peak at 
$\qord$. We display $\qord$ in panel (c) of \fref{results}.
For $J\,<\,J_{\rm{FM}}(\kappa)$ we find $q_{ord}^x =-2\pi$ which is equivalent to  $\qord = 0$.
For $J\,>\,J_{\rm{FM}}(\kappa)$, $q_{ord}^x$ increases with $J$, and it reaches the value $-4\pi/3$
at the isotropic point $J=1$. There it is independent of the value of $\kappa$ in agreement 
with Ref.~\cite{sac92}. 
For $\kappa \gtrsim 1/3$, the behaviour of $q_{ord}^x$ as a function of $J$ 
is as expected: it increases monotonously towards  $|q_{ord}^x| = \pi$ as 
$J \rightarrow \infty$. 
However, for $\kappa \lesssim 1/3$, the function $q_{ord}^x(J)$ develops 
a minimum at $J \approx 1.5$ which becomes more pronounced as $\kappa$ 
decreases, {\it i.~e.}~as the importance of quantum effects increases. 
At present we have no physical explanation for this behaviour of $\qord$.
The behaviour of the saddlepoint values of the Lagrange multipliers 
$\lambda_a$ and $\lambda_c$ as functions $J$ is displayed in panel (d) of \fref{results}
for various values of $\kappa$. As we have emphasised above, $\lambda_a \ne \lambda_c$ for 
general anisotropy, $J \ne 1$. Only at the isotropic point we find 
$\lambda_a(\kappa)= \lambda_c(\kappa)$  
in accordance with the expectation.
Again, we have no intuitive physical explanation
for the behaviour of these parameters as functions of $J$ and $\kappa$.  
Quite unexpectedly 
$\lambda_c$  drops to zero at the boundary $J_{\rm{DC}}$ of the DC phase.

In  the vicinity of the FM-IC and the IC-DC phase boundary, where $Q_1$ respectively 
$Q_2$, are small one can expand $E_{\rm{MF}}$, \eref{EMFK}, with respect to these variables 
and solve the extremum problem analytically. The expansion reads  
\be
\EMF(Q_{\alpha}) = e_{0\alpha} + r_{\alpha}\,|Q_{\alpha}|^2 +
g_{\alpha}\,|Q_{\alpha}|^4 + \mathcal{O}(|(Q_{\alpha}|^6)\,,
\label{EMF4}
\ee
where $\alpha = 1(2)$ for the FM-IC(IC-DC) phase boundary. After the stationarity conditions 
$\EMF(Q_{\alpha})$ with respect to the variables $Q_{\beta}$, $\beta=2(1)$, $\lambda_a$,
$\lambda_c$ have been solved the coefficients $r_{0\alpha}$ and $g_{\alpha}$ are known as  
functions of $J$ 
and $\kappa$. Our numerical results suggest that $g_{\alpha}>0$.   Similarly as in a 
Landau-Ginzburg expansion,  the  FM-IC and the IC-DC phase 
boundaries are then obtained as the solutions of $ r_{\alpha\,0}=0$. These solutions are 
 presented as solid lines in \fref{pd}. 
Obviously, the numerical data agree reasonably well with these analytic solutions. 
Small discrepancies arise in the numerical data because the energy surface
$\EMF(Q_1,Q_2,\la,\lc)$ is very shallow in the vicinity of the saddle 
point.
On these boundaries and inside the FM and the DC phases 
the values of $Q_1$, $Q_2$ and of the ordering wave vector remain constant as is shown 
in the corresponding panels of \fref{results}. Using a similar expansion technique for 
the \SPN{}-analog  
of the spin-spin correlation function,  we have also been able to obtain the asymptotic 
behaviour of this function on the phase boundaries analytically. We find that there is 
indeed LRO along the entire phase boundaries up to arbitarily small values of $\kappa$. 
However, while  along the FM-IC phase boundary 
there is LRO between any pair of spins $\S_{\mu\,i}$, $\S_{\nu,j}$, $\mu,\,\nu = a,\,b,\,c$,
 along the IC-DC phase boundary, 
there is LRO only between the spins on the c sites $\S_{c\,i}$, see \fref{aniskago}. 

To conclude we wish to compare the results presented here for the anisotropic \kago{} 
antiferromagnet (AF), \eref{akago}, with those obtained by Chung et al.  
\cite{chu_mar_mck01} for the \SPN{} symmetric generalisation of the anisotropic triangular AF. 
With increasing values of $J$, which controls the anisotropy in both models, we find the same   
succession of phases as in ref.~\cite{chu_mar_mck01}: a collinear phase, an incommensurate 
phase and a decoupled chain phase. However, while LRO disappears for sufficiently small 
values of $\kappa=2S$ for all values of $J$ in the triangular AF, we find LRO in the 
entire FM phase and along the 
FM-IC and the IC-DC phase boundaries up to arbitrarily small values of $\kappa$. 
For both models, the mean-field \SPN{} approach predicts a DC phase which, however, is certainly 
not described faithfully by this approach. 
Qualitative consideration of the role of fluctuations in this phase 
lead the authors of ref.~\cite{chu_mar_mck01} to the conclusion 
that there is spin Peierls order in the DC regime of 
the anisotropic triangular AF. Such a picture is still elusive for the anisotropic \kago{} AF.  
However, by applying perturbation theory to  the trimerised version of the model \eref{akago}, 
i.~e.~to a model for which 
the couplings on e.~g.~the upward pointing triangles are uniformly weaker than 
those on the downward pointing triangles \cite{mi98}, we find that for $J\gg1$ 
the dynamics of the c spins of our model, see \fref{aniskago}, is effectively described by  
the Hamiltonian of the anisotropic triangular AF with strong coupling 
in one of the three lattice directions. Therefore, according to the arguments of 
ref.~\cite{chu_mar_mck01} there should be spin Peierls order between these spins. 
Since, on account of the trimerisation, the spins on the $a$ and $b$ sites in \fref{aniskago} 
are already paired in nearest neighbour singlet pairs, one may thus 
speculate that spin Peierls order also prevails in the DC phase of the anisotropic \kago{} AF.  

\ack
One of the authors (HUE) acknowledges a useful discussion with F.~Mila.
The work at the University of Waterloo was supported by the Canada Research
Chair (Tier I, Michel Gingras).

\end{document}